\documentclass{aa}
\usepackage{graphicx}
\usepackage{txfonts}
\usepackage{color}
\usepackage{tabularx} 
\usepackage{adjustbox}
\usepackage{ulem}
\newcommand{\be}{\begin{equation}}
\newcommand{\ee}{\end{equation}}
\newcommand{\bea}{\begin{eqnarray}}
\newcommand{\eea}{\end{eqnarray}}

\begin{document}

\title{Observations of solar chromospheric oscillations at 3 mm
with ALMA
}
\author{S. Patsourakos \inst{1}
\and C.E. Alissandrakis\inst{1}
\and A. Nindos \inst{1}
\and T. S. Bastian\inst{2}}
\institute{Physics Department, University of Ioannina, Ioannina GR-45110,
Greece\\
\email{spatsour@uoi.gr}
\and
National Radio Astronomy Observatory, 520 Edgemont Road, Charlottesville VA
22903, USA}

\date{Received: Accepted:}

 
  \abstract
 {}
{To study spatially-resolved chromospheric oscillations of the quiet Sun in the mm-domain at a resolution of a few arcsec, typically 2.4\arcsec$\times$4.5\arcsec.}
{We used Atacama Large millimeter and
sub-millimeter Array (ALMA) time-series
of interferometric observations  of the quiet Sun 
obtained at 3 mm
with a 2-s cadence and a spatial resolution of a few arcsec.
The observations were performed on March 16, 2017 and seven 80\arcsec$\times$80\arcsec 
fields-of-view going from disk center to limb were covered, each one observed for 10\,min, 
therefore limiting the frequency resolution of the power spectra to 1.7 mHz.
For each field of view, masks for cell and network were derived, and 
the averaged power spectral densities 
(PSDs)
for the entire field of view, cell and network were computed.
The resulting power spectra were fitted with an analytical function in order to derive the frequency and the root-mean-square (rms) power
associated with the peaks.
The same analysis, over the same fields of view and
for the same intervals, was performed for simultaneous
Atmospheric Imaging Assembly (AIA)  
image sequences in 1600 \AA.}
{Spatially-resolved chromospheric oscillations at 3 mm,
with frequencies of $ 4.2 \pm 1.7$\,mHz 
are observed in the quiet Sun, in both cell and network.
The coherence length-scale  of the oscillations is commensurate with
the spatial resolution of our ALMA observations.
Brightness-temperature fluctuations in individual pixels could reach up to a few hundred K,
while the spatially averaged power spectral densities 
yield rms in the range $\approx$ 55-75 K,
i.e., up to $\approx$ 1 $\%$ of the averaged
brightness temperatures and exhibit a moderate increase towards the limb.
For AIA 1600 \AA \, the 
oscillation 
frequency is $3.7\pm1.7$ mHz.
The relative  rms is up to 6\% of the background intensity, with a
weak increase towards disk center (cell, average).
ALMA 3 mm time-series lag AIA 1600 \AA \, by $\approx$ 100 s, which corresponds
to a formation-height difference of $\approx$ 1200 km, representing a novel determination
of this important parameter.}
{The ALMA oscillations 
that we detected
exhibit higher amplitudes than
those derived from
the lower ($\approx$ 10\arcsec) resolution
observations at 3.5 mm 
 by White et al. (2006). Chromospheric oscillations
are, therefore, not fully resolved at the length-scale of the chromospheric network, and possibly not
even at the spatial resolution of our ALMA observations. 
Any study of transient brightenings in the mm-domain should take into account
the oscillations.}


   \maketitle
%

\section{Introduction}
Oscillation and wave phenomena are 
ubiquitous throughout the solar atmosphere (see the reviews by Tsiropoula et al. 2012 and
Jess et al. 2012 and the references therein). 
Chromospheric oscillations with periods between 3-5 minutes could represent the intrusions of the
photospheric p-modes (Leighton et al. 1962) into the chromosphere
(e.g., Jefferies et al. 2006). However, given
the inhomogeneous nature of the chromosphere, along
with the transition between plasma-dominated dynamics (i.e., high plasma $\beta$) in the photosphere
to magnetic-dominated dynamics (i.e., low plasma  $\beta$) in the chromosphere,
a complex picture of wave and oscillation phenomena has been established for the chromosphere which includes mode conversions, reflections, interferences, and 
shocks.
(e.g., Wedemeyer-B{\"o}hm et al. 2009).
The importance of chomospheric oscillations and waves is manifold as they 
could supply a means to heat plasmas and  a probe of atmospheric conditions, including
the magnetic field.

Almost all observations of chromospheric oscillations in either intensity or Doppler shifts
have been performed in the visible or the extreme ultraviolet (EUV).
The derivation of physical parameters from such observations,
relies on complex diagnostics with complicated physical effects such as departures from local thermodynamic
equilibrium, partial and time-dependent ionization being highly relevant (e.g., Leenaarts et al. 2013).

On the other hand, observations of free-free emissions in the radio domain are devoid of the
complications described above, and in addition, as per Rayleigh-Jeans law, the observed
brightness temperatures are directly (linearly) linked to the plasma temperature 
(e.g., Shibasaki et al. 2011). 
Given now the standard variation of temperature and density with height in the solar atmosphere,
the chromosphere is probed in mm-wavelengths by the above mechanism. 

\begin{figure*}
\centering
\includegraphics[width=0.8\textwidth]{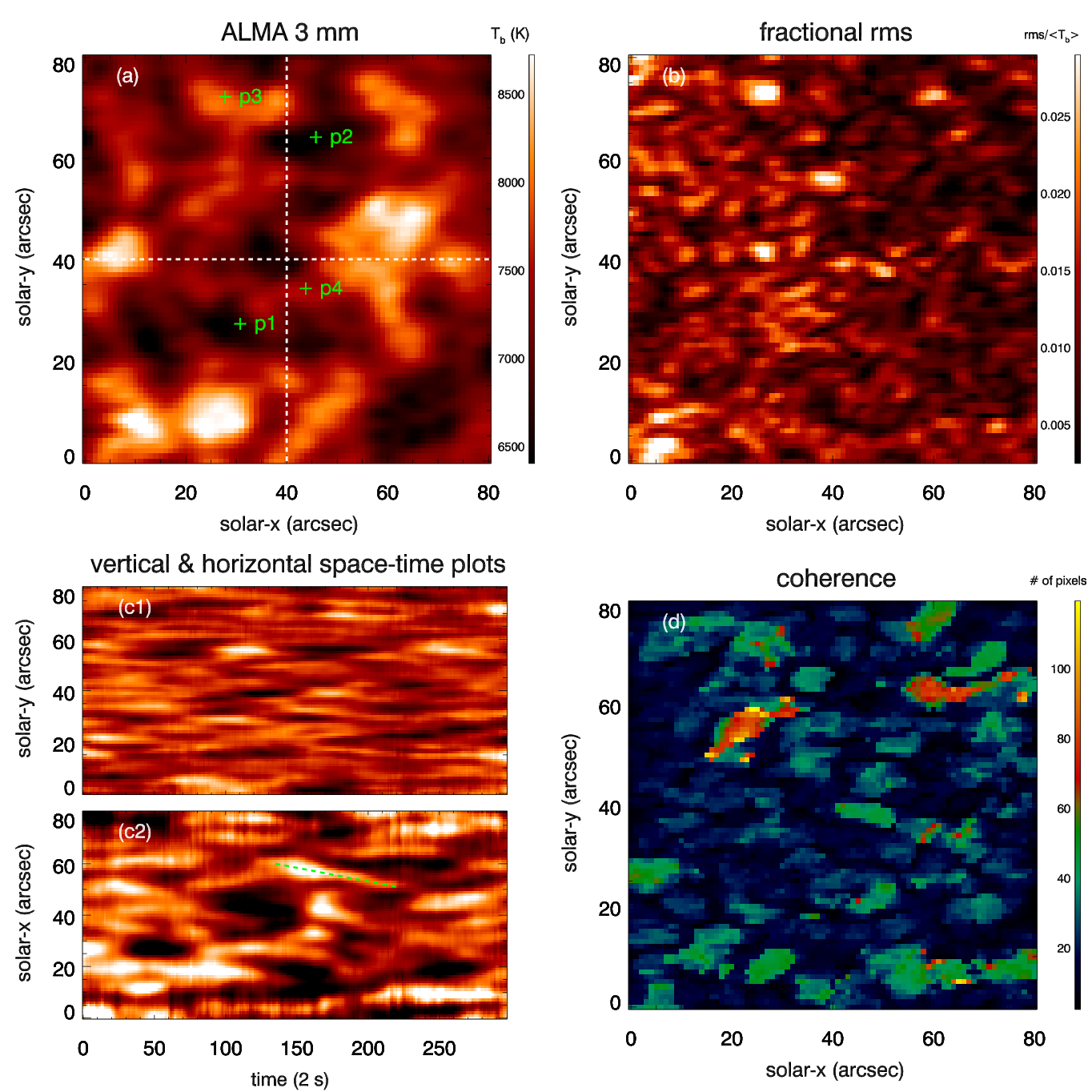}
\caption{Summary plots for target 5. (a): temporally-averaged
ALMA 3-mm image, 
(with p1-p4 we show the four pixels used in Figure \ref{lc}); (b): fractional rms of T$_b$; (c):
space-time plots of $T_{b}$ corresponding to a vertical (upper panel) and horizontal
(lower panel) cut through the center of image of (a); from each light-curve its 
temporal average was subtracted and the color-scale was saturated to differences
of $\pm$ 200 K: (due to the non-circular ALMA
beam the resolution in the vertical direction
is inferior to that in the horizontal direction: see
Table-1 of Nindos et al. 2018); (d) coherence map displaying the number of pixels in the neighborhood
of each pixel  with linear correlation coefficient of the corresponding
light-curves of at least 0.7. Images are oriented with celestial north up.}
\label{summary}
\end{figure*}

Unfortunately, even moderate spatial resolution observations of the solar chromosphere in the mm-domain 
are rather scant. A notable exception are the BIMA observations
of White et al. (2006) and  Loukitcheva et al. (2006). Their BIMA observations at 3.5 mm
covered both quiet Sun (QS)
and active region targets with fields of view truncated to $72$\arcsec.
The spatial resolution of $\approx$ 10\arcsec\
of the BIMA observations did not allow
a complete separation
 between network and 
cell elements, and the above authors had to use a Ca\,{\sc ii} spectroheliogram to achieve this task.
Therefore, some mixing between cell and network elements was unavoidable. 
White et al. (2006) reported  p-mode
oscillations
with periods of 3-5 minutes, and even longer period oscillations for the case of the
network.  

The commencement of solar observations with ALMA in 2015 marked a
new and exciting era of chromospheric investigations. This is because of the superior
spatial resolution and  sensitivity that ALMA is offering.
In particular, the interferometric
ALMA observations reach spatial resolution of a few arcsec or less, 
a significant improvement over the 
previous observation in the mm-domain, which indeed starts  to somehow bridge
the gap with commensurate observations in the infrared, optical and EUV wavelengths (for a review of the ALMA capabilities
in solar observing  see Wedemeyer et al. 2016). Observations of the quiet Sun
with ALMA include Alissandrakis et al. (2017);  Bastian, et al. (2017);
Shimojo et al. (2017);
White et al. (2017);
Braj${\v{s}}$a et al. (2018);
Nindos et al. (2018); Yokoyama et al. (2018);
Jafarzadeh et al.(2019); Loukitcheva et al. (2019);
Molnar et al. (2019); 
Selhorst, et al. (2019).

We hereby present the first ALMA observations of quiet Sun chromospheric oscillations at  3 mm.
Our observations allow for the first time to study chromospheric oscillations in the mm-domain with
a spatial resolution adequate to clearly distinguish between cell and network and, in addition, to
follow the center-to-limb variation   of their properties, including 
their frequency and relative amplitude.
Moreover, we compare with AIA (Lemen et al. 2011) observations taken in its 1600 \AA \, channel, and derive
temporal lags between the ALMA and AIA oscillations. 
In Section 2 we discuss our observations and their analysis; Section 3 contains the results of our analysis
and we conclude in Section 4 with a summary and discussion of our results.

\begin{figure*}
\centering
\includegraphics[width=0.8\textwidth]{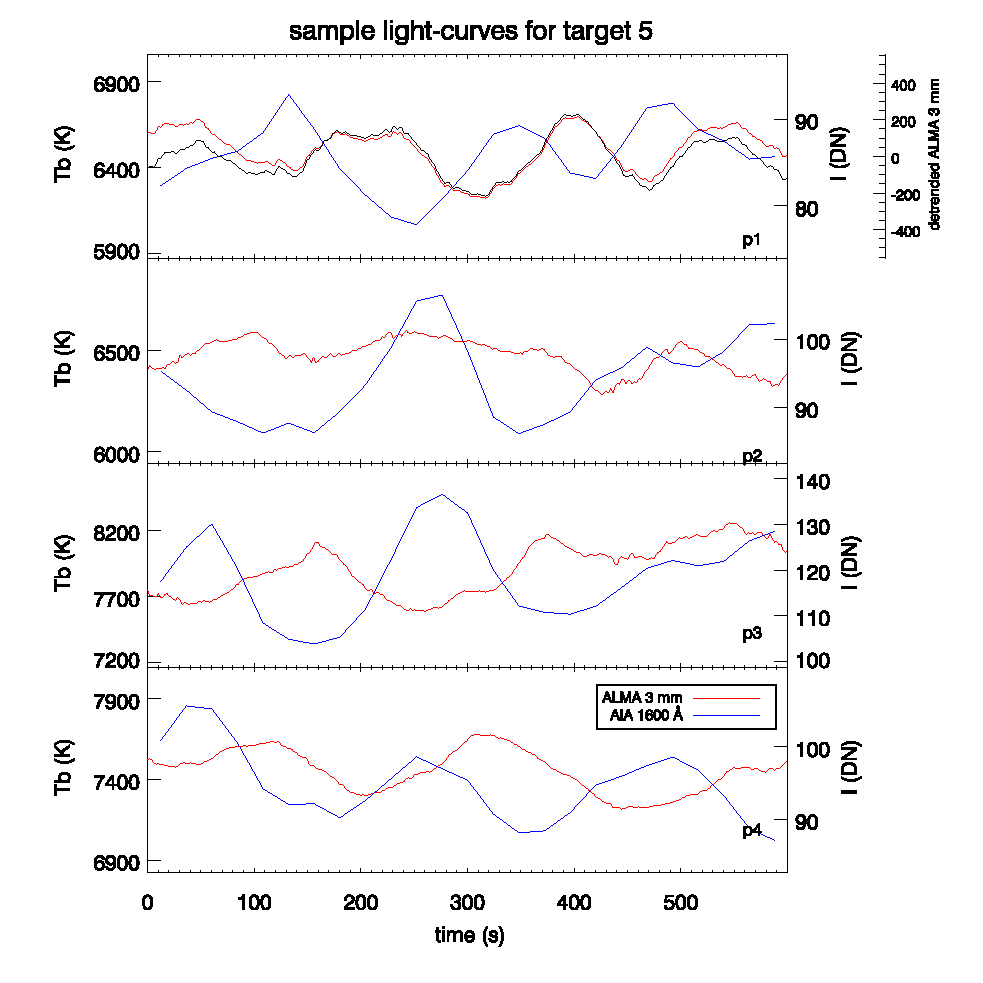}
\caption{ALMA 3 mm  (red curve) and AIA 1600 \AA \, (blue curve) light-curves for the 
four selected pixels shown in Figure \ref{summary}. The 
black curve of the upper 
panel 
shows the ALMA light curve after the subtraction of a 3rd-degree polynomial fit.}
\label{lc}
\end{figure*}

\section{Observations and data analysis}

We used  ALMA observations of the quiet Sun described by Nindos et al. (2018).
In short, seven  120$^{"}$ circular fields of view (targets) were observed at 3 mm (100 GHz) on  March 16, 2017.
The ALMA targets, numbered from 1 to 7,
correspond to $\mu$=[0.16, 0.34, 0.52, 0.72, 0.82, 0.92, 1] along
a line at a position angle of 135\degr,
therefore supplying a center-to-limb coverage (see Table 1 of Nindos et al., 2018). Since the field of view (FoV) of target 1 
included
off-limb locations, we refrained from using it for further analysis, therefore limiting
ourselves to targets 2-7. Each target was observed for 10 minutes with a 2-s cadence.
The resulting ALMA images have  1\arcsec\ 
pixels; the 
spatial resolution, as resulting from the beam size, for all targets was 
2.4\arcsec$\times$4.5\arcsec,
with the exception of target 7 
where the resolution was 
 2.3\arcsec$\times$8.1\arcsec.
The ALMA images were already corrected for the (small) effect of solar rotation.
More details regarding the reduction of the ALMA data  can be found
in Nindos et al. (2018). 

We extracted a 80\arcsec$\times$80\arcsec region from the original 120\arcsec\ diameter FoV at   
the center of each target to avoid artifacts resulting from the primary beam correction towards the
edges of each FoV.
Given that the mean diameter of supergranulation
cells is around 20-30\arcsec, our truncated FoVs encompass a significant
number of supergranulation cells to allow  for the determination of meaningful statistics of the quiet
Sun.

We also analyzed AIA images
at 1600 and 304 \AA, convolved with the ALMA resolution, both for purpose 
of comparison and as a check of our 3 mm computations. 
Note here that
Howe et al. (2012), from the analysis of  QS oscillations in the 1600 \AA \, and 1700 \AA \,
channels of AIA, showed that the two emissions are formed at similar heights, which suggests
that the 1600 \AA \, channel emission is mainly of chromospheric (i.e., continuum) and not transition region
(i.e., C IV line) origin.  

We selected the same regions and 
the same time intervals used for the ALMA analysis. The main difference is that the cadence of the AIA images 
is lower, 12 s and 24 s for 304 \AA \, and 1600 \AA \, respectively. 
This limits the highest PSD frequency to 42 mHz and 24 mHz 
respectively, compared to 250 mHz of the ALMA data set; all three values are 
though
well above the p-mode frequency of $\sim$ 3 mHz. 
The frequency resolution of all three data sets, determined by the duration of the time series, 
is 1.7 mHz. The AIA images were first corrected for differential rotation.
Next, the AIA and ALMA were co-aligned by first considering the 
celestial orientation of the ALMA images 
and then cross-correlating the corresponding temporally-averaged images for each target.

An example of our ALMA observations can be found in Fig. \ref{summary}, and refers
to target 5 ($\mu=0.82$).
In panel (a) we display the temporally-averaged ALMA 3 mm image over the corresponding  10-min interval.
Thanks to the spatial resolution of our ALMA observations,
the differentiation between cell and network is obvious
and several network elements could be readily discerned. In panel (b) of Fig. \ref{summary}  we display a fractional rms map
which contains the fraction of 
the rms of the light-curve at each pixel
with the corresponding temporal average of  $T_{b}$ ($<Tb>$). 
Prior to the calculation of the rms for each light curve, a 3rd-degree polynomial fitting
was subtracted.  
Weak fluctuations of $T_{b}$ not exceeding few percent can be observed throughout the FoV; the mean value is around 1 $\%$.
We  are not though able  to observe a significant correspondence between
the temporally-averaged image of panel (a) and the fractional variability map of panel (b) (linear correlation
coefficient of 0.12).

Panel (c) contains space-time plots
of $T_{b}$ corresponding to a vertical (upper plot) and a horizontal (lower plot) cut through the center of the image of panel (a).
To enhance visibility of oscillatory behavior we subtracted the corresponding temporal average from the light curve at 
each location along the cuts. $T_{b}$-oscillations correspond to alternating black and white patches along the time axis at a given location.
The amplitude of the oscillations at individual pixels could reach 
values of $\approx$ 350 K.
Such patterns are common and  cover both cell and network locations. One to two periods are observed. Patches
showing oscillatory behavior 
span 
several pixels along the 
spatial direction 
and
correspond to scales of up  $\approx$ 10\,\arcsec. This  suggests that the oscillations exhibit a degree of spatial coherence.
We also note the occasional appearance of bright lanes, inclined with respect to the spatial axis
(e.g., corresponding to the dashed green line in the lower plot of panel (c)).
Such lanes are indicative of an apparent motion of a $T_{b}$ front. The analysis of such fronts deserves a separate study. 

In addition, in Figure \ref{lc} we supply sample light-curves for four
pixels
for target 5, indicated as p1-p4 in Figure 1a,
with the first two corresponding to cell and the other
two to network. Again, the oscillatory behavior is rather obvious.
Indeed p1 shows higher frequency oscillations. 

\begin{figure*}[]
\centering
\includegraphics[width=0.8\textwidth]{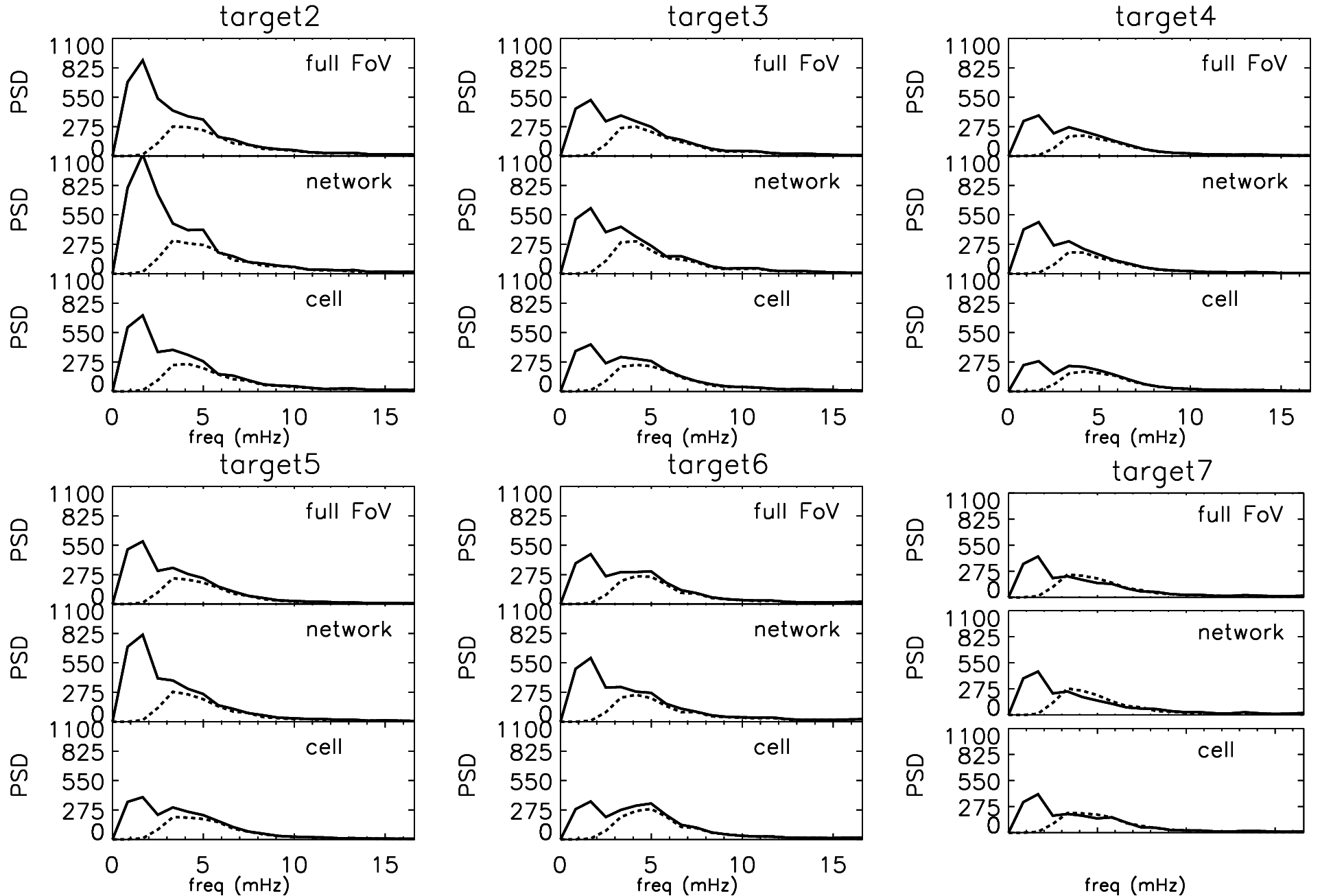}
\caption{Power spectra from full FoV (first row), network (second row) and cell (third row) for
targets 2-7. Solid  lines correspond to spectra
computed after the subtraction of the average 
from the light-curves of individual pixels, dashed lines show the spectra computed 
after the subtraction of a 3rd-degree polynomial fit.
All  spectra have been scaled by a factor equal to 4815 $\mathrm{{K}^2/Hz}$.}
\label{psdcomp}
\end{figure*}

We then investigated in more detail the spatial coherence 
of the $T_{b}$ fluctuations as follows.
We first calculated the cross-correlation coefficient between  
the light curve of each pixel of the FoV and the light curves of the remaining pixels.
Prior to  the calculation of the cross correlation we subtracted a 3rd-degree polynomial fitting from each light curve.
Next, initiating a region-growing algorithm at each pixel, we determined clusters of
pixels with  cross-correlation coefficient 
above 0.7 and counted the number of pixels per cluster. This pixel count
is displayed in  the coherence map of panel (d) of Fig. 1.
Adjacent pixels with high correlations of their light-curves imply
coherent (in-phase) $T_b$ fluctuations (oscillations). Again, from the inspection of the fractional variability map of panel (b),
we cannot observe any strong correspondence between the coherence
map and the temporally-averaged image of panel (a) (linear correlation coefficient of 0.1).
The average number  of high-correlation pixels in the vicinity of a given pixel in the FoV is $\approx$ 20.
This corresponds to a square with a side of $\approx$ 4\arcsec, which is comparable with
the ALMA spatial resolution. Therefore, this sets an upper limit on the spatial coherence of QS fluctuations we observe.
 This number exhibited  a small variation from 18-25 for all considered targets, with the exception of target 7 for
which it was equal to 45, apparently due to the lower resolution of the ALMA images for this target. 
Panels (b) and (d) of Fig. 1 exhibit 
some similarities; this comes to no surprise
since both supply different and complementary means to look into the fluctuations in the light-curves. 

Note here that 
inspection of 2D maps of the PSDs integrated over selected frequency ranges (e.g.,
covering the p-mode domain) do not reveal any correspondence with structures observed in
the original ALMA images, at variance of structures such as the "shadows" and "halos"
observed in relation to chromospheric oscillations in other wavelengths (e.g., review of Tsiropoula et al. 2012).
This lack  is possibly due to the small duration of our ALMA observations per target (i.e., $\approx$ 10 min).

\begin{figure*}[]
\centering
\includegraphics[width=0.8\textwidth,height=20cm]{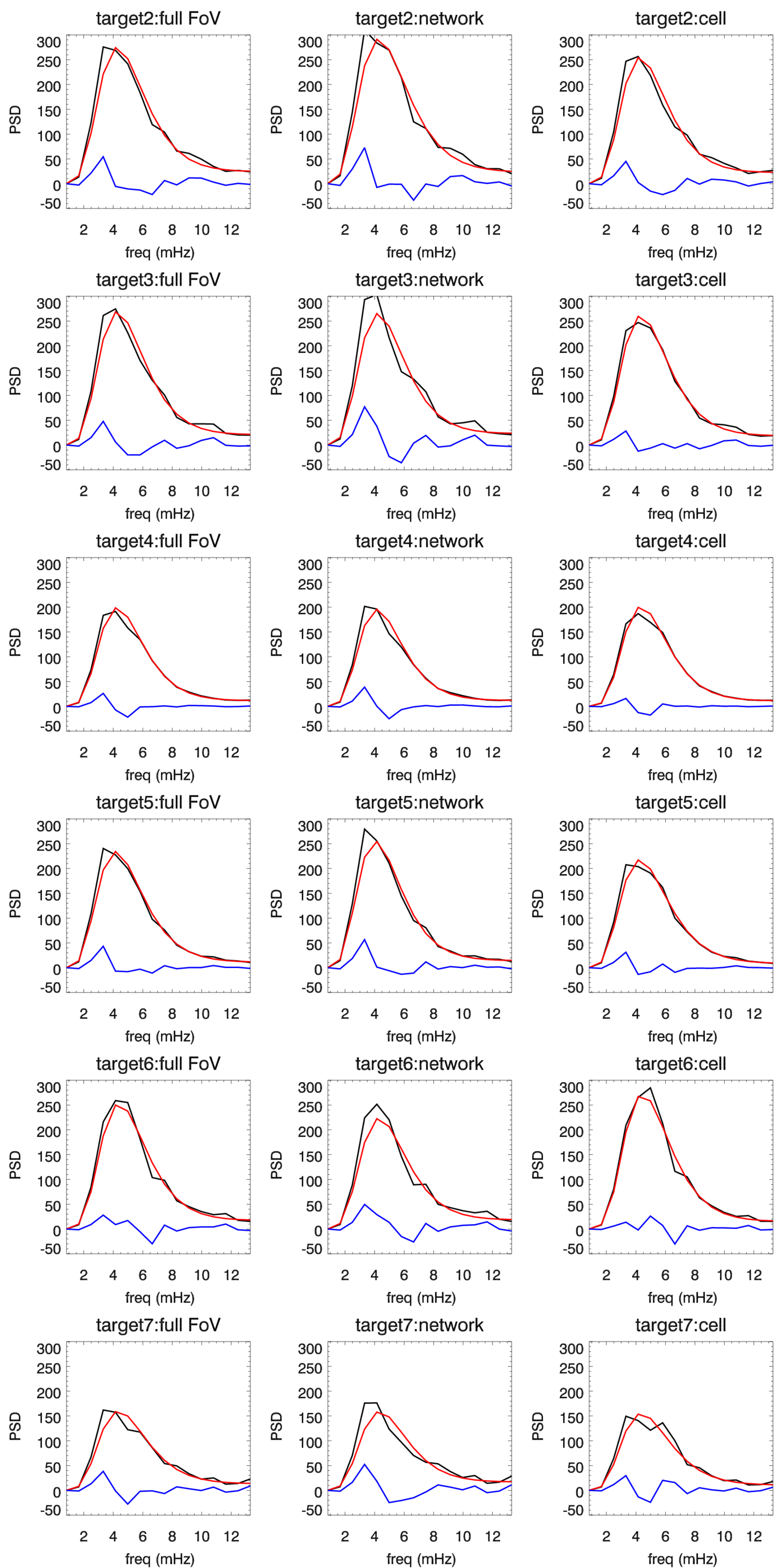}
\caption{Averaged ALMA 3-mm PSDs (black  line) for the entire FoV, network and cell (first, second and third column, respectively)
for targets 2-7 (top to bottom row); the corresponding fittings with  Equation
\ref{fitpsd} are plotted with a red line and the residuals (observed PSDs - fittings) with a blue line. All 
PSDs have been scaled by a factor equal to 4815 $\mathrm{{K}^2/Hz}$.}
\label{psdfit}
\end{figure*}

\begin{figure*}[]
\centering
\includegraphics[width=0.8\textwidth]{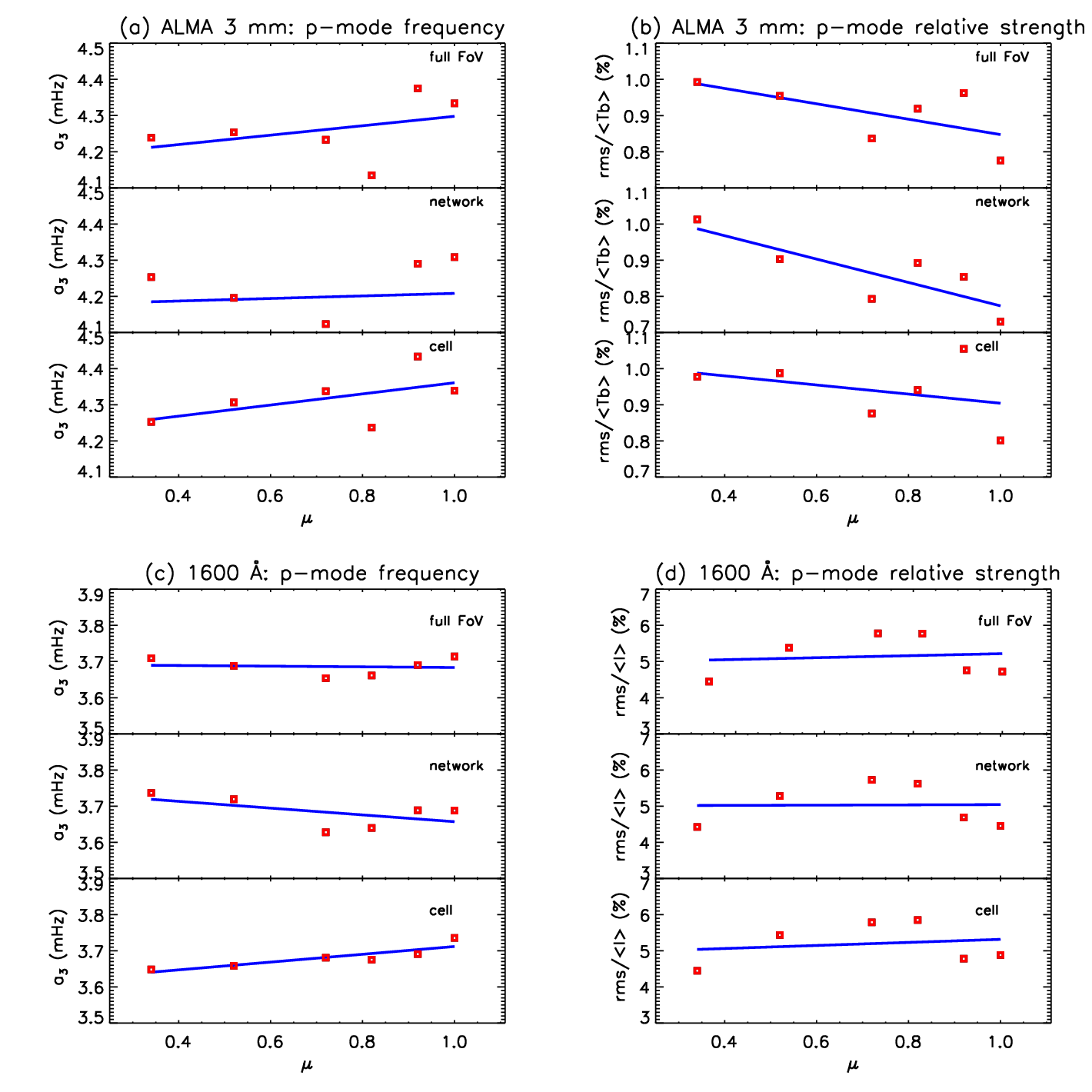}
\caption{ALMA 3 mm center-to-limb variation (red boxes) for: 
(a) p-mode frequency (i.e., $a_3$ in Equation 2),
(b) p-mode  relative strength, i.e., rms associated with p-mode/<Tb> ($\%$).
AIA 1600 \AA \, center-to-limb variation (red boxes) for: 
(c)  p-mode frequency (i.e., $a_3$ in Equation 2), (d) p-mode  relative strength, i.e., rms associated with p-mode/$<I>$ ($\%$).
Blue lines correspond to linear fits of the 
corresponding measurements.  
In each panel,  top, 
middle and bottom plots correspond to the entire FoV, network and cell, respectively.}
\label{clvalma}
\end{figure*}

\begin{figure*}[]
\centering
\includegraphics[width=0.8\textwidth,height=20cm]{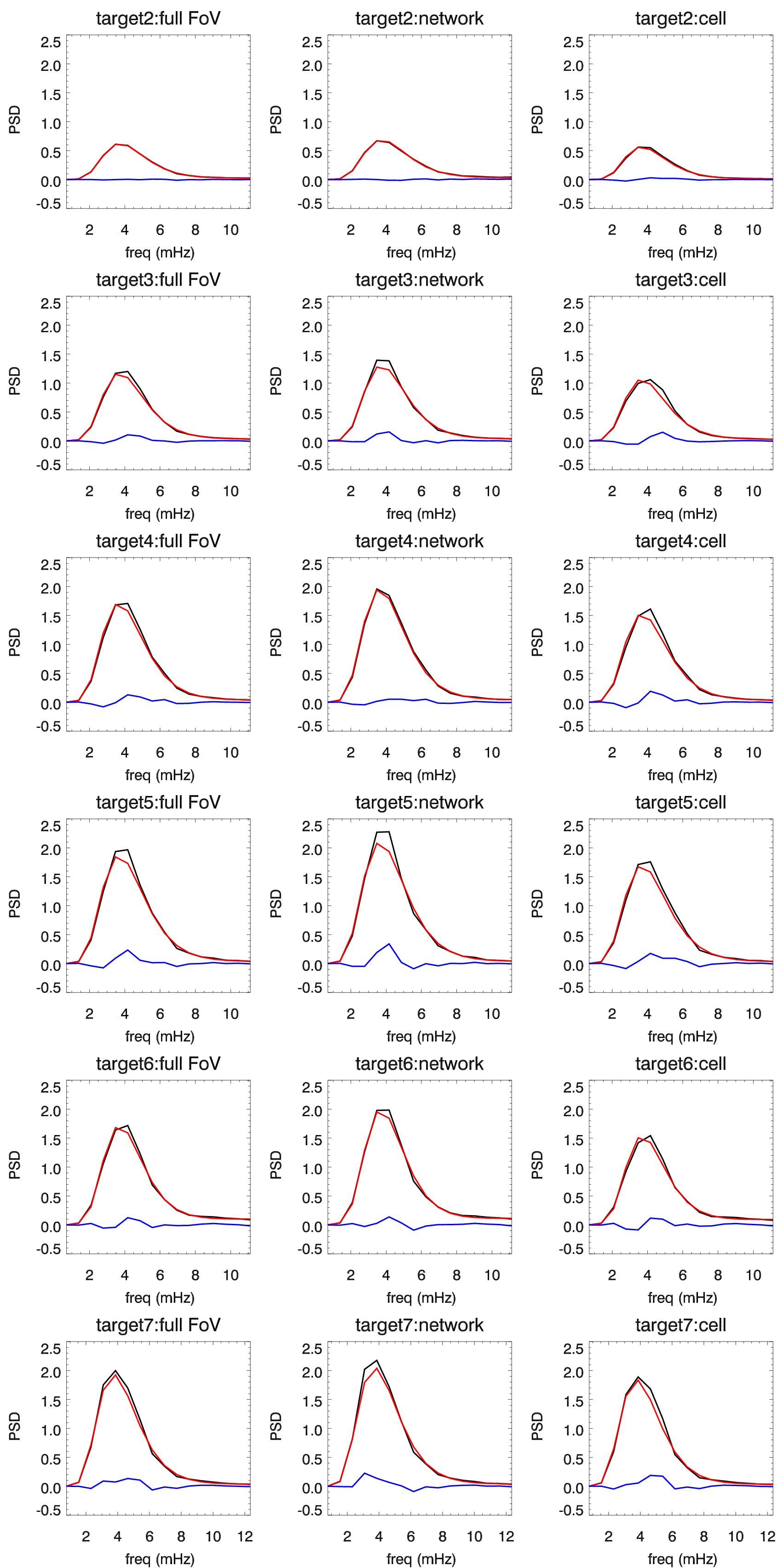}
\caption{Averaged AIA 1600 \AA \,  PSDs (black  line) for the entire FoV, network and cell (first, second and third column)
for targets 2-7 (top to bottom row); the corresponding fittings with  Equation
\ref{fitpsd} are plotted with a red line and the residuals (observed PSDs - fittings) with a blue line. All 
PSDs have been scaled by a factor equal to 5760 $\mathrm{{DN/s}^2/Hz}$;
DN correspond to the digital numbers recorded by AIA.}
\label{psdfit1600}
\end{figure*}

In order to determine the characteristics 
(e.g., amplitude, frequency)
of  the p-mode oscillations we 
performed a series of steps described below.

\subsection{Cell-Network Segregation} 
After building the temporal average of all images per target, using  the truncated-FoV
images discussed above,
we submitted the resulting average images to the masking
procedure described in Nindos et al. (2018). This essentially
calculates a low-order (i.e., second-degree) 2D polynomial fitting
of the average image, which is meant to represent the large-scale
structure of the corresponding scene. Pixels with values smaller (larger)
than these of the mask are tagged as cell (network) pixels.
As also discussed in Nindos et al. (2018), our employed method of separating
network-cell is by no means unique; it though leads
to satisfactory results as judged by visual inspection of the resulting
masks and the actual average images. 

\subsection{Calculation of Power Spectral Density}
Next, the light-curves of each pixel were submitted to a 
Fourier
analysis to derive the corresponding Power Spectral Density (PSD). 
Having $N$ evenly sampled points in a light-curve $x_{n}$ $(n=1,2,..,N)$ the corresponding
PSD is given by:
\begin{equation}
\mathrm{PSD}(f_j)=\frac{2N}{fs}{|\tilde{x_j}|}^2,
\label{psd}
\end{equation}
with $j=1,2,..,N/2$,
$\tilde{x}$ is the discrete Fourier transform of 
$x_{n}$ and $fs$ is equal to the sampling frequency.
In our application, $x_{n}$ corresponds to the observed $T_{b}$
time-series.

From each light curve we first subtracted
a 3rd-degree polynomial fitting to remove long-term trends from the data (see also White et al. 2006).
This essentially leads to 
zero mean value of the resulting light curves and therefore
quenches the zero frequency (DC term) in the corresponding power spectra. 
Moreover, it eliminates the non-periodic,  slow-varying background from the light curves,
leaving only temporal fluctuations around the mean level. Therefore,
each light curve will contribute to the  spatially-averaged PSDs according only
to the amplitude of the corresponding fluctuations and not of the intrinsic 
brightness of the spatial pixel under consideration. An example
of this procedure is given in the upper panel of Fig. 2, where the red curve
corresponds to the ALMA 3 mm light-curve for pixel p1 of Fig. 1c,
while the black curve corresponds
to the result of the subtraction of the 3rd degree polynomial fitting from the red curve.

We then calculated the PSDs
for every pixel in the FoV and averaged over
the entire FoV,
as well as the cell and network
using the masks described above.
Averaging PSDs of individual pixels, 
rather than first calculating spatially-averaged light curves and then
calculating the corresponding PSDs, guarantees that oscillatory patterns in pixels which are out-of-phase
are not smeared out. The PSDs
both before and after the subtraction of the 3rd degree polynomial fittings
are displayed in Fig. 3  and will be discussed in detail 
in Section 3.

\subsection{Fitting of spatially averaged power spectra} 
The spatially averaged PSDs 
were then fitted with the following function:
\begin{equation}
\mathrm{PSD}(f)=a_{0}+a_{1}f+a_{2}\,\exp\left(-\frac{{{(\ln f-\ln a_{3})}}^{2}}{2{a_{4}}^2}\right),
\label{fitpsd}
\end{equation}
with  $a_{0}-a_{4}$ being the  parameters of the fitting function.
The fitting function of Equation \ref{fitpsd} treats the observed PSDs as a sum
of a linear function describing the background (noise) spectrum described by the terms
$a_{0}$ and $a_{1}$, and 
with a
log-normal (Gaussian of $\ln f$), describing the peak associated with
chromospheric oscillations;
$a_2$ is the oscillation amplitude, $a_3$ 
is the frequency at maximum power and $a_4$ 
is related to the width of the spectral peak.
We note that both the amplitude and the width are determined 
by the frequency resolution 
of the time series rather than the  properties of the p-modes, 
thus the physically meaningful quantities 
are the frequency at maximum power and the integral of the exponential term over frequency, which evaluates to
\begin{equation}
a_2\int_0^\infty \exp\left(-\frac{{{(\ln f-\ln a_{3})}}^{2}}{2{a_{4}}^2}\right) \,df
=a_2\sqrt{{2\pi}}a_4   \exp \left(\frac{a_4^2+2\ln\left(a_3\right)}{2} \right)
\label{rms}
\end{equation}
This represents the total power of the oscillations and is equal to the rms of their amplitude.

The choice of Equation \ref{fitpsd} to fit the PSDs is by no means unique; 
however, as can be seen in Figs. 4 and 6, displaying the 
averaged PSDs for ALMA 3 mm and AIA
1600 \AA\,,respectively, it gives
a good representation of the observed PSDs,
and their  peaks in particular, as required for our purposes
to characterize the observed chromospheric oscillations. Log-normal fittings of peaks in  PSDs
have been employed in other investigations of oscillatory phenomena (e.g., Ireland et al. 2015; Morton
et al. 2019). Regarding the choice of a linear function  to describe  the noise in the PSDs, hence
implying a white-noise spectrum, it is well-known that the noise in solar PSDs
could be represented by power-laws on several occasions/features (e.g., Ireland et al, 2015; Auch{\`e}re et al. 2016).
However, the short duration of the analyzed sequences per target (i.e., $\sim$ 10 min),
does not allow to extend the PSDs to significantly lower frequencies than these
corresponding to the p-mode regime so 
as to obtain a fair representation of the noise
spectrum at both sides of the corresponding peaks. In addition, since we focus on narrow frequency windows
around the peaks, a linear background is deemed as a reasonable choice.
A final note regarding this point is in order: as  can be seen by simple inspection
of Figs. 4 and 6
the p-mode peaks stand significantly above the corresponding backgrounds. 
Therefore, it is reasonable to expect that the specifics of the employed background model   
will  have a minor impact on the characteristics of the spectral peak fittings (e.g., amplitude).
The fittings were performed in the frequency range
of $\approx$  [0.008, 13.3] mHz, spanning essentially the spectral peaks
and the adjacent background
 seen in
Figs. 4 and 6.

\begin{table}[t]
\caption{ALMA 3 mm: center-to-limb variation variation of oscillation parameters.}
\begin{tabular}{l|cccccc}
\hline 
             &\multicolumn{6}{c}{$\mu$}                \\
             & 1.00 & 0.92 & 0.82 & 0.72 & 0.52 & 0.34 \\
\hline 
&\multicolumn{6}{c}{Full FoV}\\
$a_{3}$ (mHz)& 4.3 &  4.3 & 4.1 &  4.2 &  4.2 & 4.2 \\
rms (K)      &  56 &  70 &  67 &  61 &  71 &  72 \\
rms/$<T_b>$    & 0.78  & 0.96  & 0.92  & 0.84   & 0.95  & 0.99   \\
&\multicolumn{6}{c}{Network}\\
$a_{3}$ (mHz)&  4.3 & 4.2 &  4.0 &  4.1 &  4.2 & 4.2  \\
rms (K)      &  55 &  65  & 68 &  60 & 70  & 77 \\
rms/$<T_b>$    & 0.73 &   0.85& 0.89 &  0.79 &  0.90 &  1.01 \\
&\multicolumn{6}{c}{Cell}\\
$a_{3}$ (mHz)&   4.3 & 4.4  &  4.2 &  4.3 &  4.3 &  4.2 \\
rms (K)      &  55 & 73  &  66 &  61 &  71 &  69 \\
rms/$<T_b>$    &  0.80 &  1.05  & 0.94  & 0.88  &  0.99 & 0.98 \\
\hline 
\label{tablealma}
\end{tabular}
\end{table}

\section{Results}
\subsection{ALMA 3 mm}
The spatially-averaged PSDs for the entire FoV, network and 
cell for all  targets are displayed in Fig. \ref{psdcomp}.
In addition to displaying PSDs resulting from the subtraction of a 3rd degree polynomial fitting from
the original light curves (dashed lines in Fig. \ref{psdcomp}; $\mathrm{PSD_{poly}}$) as discussed above,
we are also displaying PSDs ($\mathrm{PSD_{av}}$) resulting from the subtraction of the light-curve average 
$\mathrm{PSD_{av}}$)  instead
(solid lines in Fig. \ref{psdcomp}). We first note that
only the $\mathrm{PSD_{av}}$ curves  exhibit a low-frequency
peak close to 0 corresponding to long-term trends
in the original light curves; with an increase towards the limb.
Such peaks do not
show up in the $\mathrm{PSD_{poly}}$ curves as a result of the application
of the polynomial fitting to the original light curves.
On the other hand, both  $\mathrm{PSD_{poly}}$ and
$\mathrm{PSD_{av}}$ 
exhibit well-defined strong peaks in the p-mode range, i.e., $\approx$ 3-4 mHz with similar 
locations and amplitudes. Therefore, the identification of 
p-modes in our ALMA observations 
is robust. This conclusion,
along   with the  discussed-above low-frequency peak
in the $\mathrm{PSD_{av}}$ curves, which prevents  
a proper mapping of the p-mode peaks,
led us to use the $\mathrm{PSD_{poly}}$ curves for further analysis.
Note here that experimentation with different degree polynomials
(i.e., 1-4) showed that the higher the degree of the polynomial, the smaller
the amplitude of the peak around the p-mode domain. Therefore,
the derived p-mode rms amplitudes from the use of a 3rd degree polynomial
fitting of the original light-curves
could be seen as upper limits.
Obviously, employing longer duration ALMA time-series will remedy
some of the issues we encountered here.
Finally, for frequencies above the p-mode range both  $\mathrm{PSD_{poly}}$ and
$\mathrm{PSD_{av}}$ are practically the same.
Note here that although the duration of the time series is barely equal to twice the period of p-mode oscillations, 
the peak stands out clearly in the power spectra.

\begin{table}[t]
\caption{AIA 1600 \AA\,: center-to-limb variation variation
of oscillation parameters.}
\begin{tabular}{l|cccccc}
\hline 
             &\multicolumn{6}{c}{$\mu$}                \\
             & 1.00 & 0.92 & 0.82 & 0.72 & 0.52 & 0.34 \\
\hline 
&\multicolumn{6}{c}{Full FoV}\\
$a_{3}$ (mHz)&  3.7 & 3.6  & 3.6 &  3.6 & 3.6 &  3.7 \\
rms (DN)      &  5.7  & 5.3  &  5.8 & 5.5  & 4.6  & 3.3  \\
rms/$<I>$    & 4.7  & 4.8  & 5.8  &  5.9 & 5.4  & 4.5 \\
&\multicolumn{6}{c}{Network}\\
$a_{3}$ (mHz)&  3.6 &  3.6 &  3.6 &  3.6 & 3.7  & 3.7  \\
rms (DN)      &   5.9&  5.7 &  6.2 & 5.9 & 4.8  &  3.5 \\
rms/$<I>$    & 4.5  &  4.7 &  5.6 &  5.7 &  5.3 & 4.4  \\
&\multicolumn{6}{c}{Cell}\\
$a_{3}$ (mHz)& 3.7  & 3.6  & 3.6  &  3.68 & 3.6  &  3.6 \\
rms (DN)      &   5.5&  5.0 & 5.6  &  5.2 & 4.3  &  3.1 \\
rms/$<I>$    & 4.9  & 4.8  & 5.9  &  5.8 & 5.4  & 4.5 \\
\hline 
\label{tableaia}
\end{tabular}
\end{table} 

The resulting averaged PSDs 
for the full FoV, cell and network
were then fitted with the function 
of Equation \ref{fitpsd}. From Fig. \ref{psdfit} we observe that the employed
fitting function (red lines) does a good job at reproducing the observed PSDs (black lines) with
residuals (blue lines) not exceeding $\approx 15 \%$. 

We finally plot in Fig. \ref{clvalma}
the center-to-limb variation of the  peak frequency (i.e.,
parameter $a_{3}$ of Equation \ref{fitpsd}; panel a)
and the relative ($\%$) rms of the p-modes     
(i.e., the ratio between
the rms corresponding to the Gaussian part of Equation \ref{fitpsd}
as resulting from application of Equation \ref{rms}
and
the average $T_{b}$)
for the entire FoV, cell and network.
The corresponding values are tabulated
in Table 1.  From Fig.  \ref{clvalma} we observe that the
oscillation frequencies are $\approx$ 4.2 mHz (the moderate apparent increase
towards the limb is within the frequency resolution); the oscillation rms is small
($\approx$ 55-75 K), and conversely up to $\approx$ 1 $\%$ of the averaged
$T_{b}$. It exhibits a moderate increase towards the limb.
The peak frequency of the oscillations is somewhat larger in the cell
compared to the network. 
We note, however, that frequency differences are considerably smaller 
than the 1.7\,mHz  frequency resolution of the PSD.

\subsection{AIA 1600 and 304 \AA}
We degraded our AIA 1600 \AA\ image sequences to the ALMA spatial resolution and submitted them to the same
analysis as the 
ALMA data, discussed previously.
p-mode oscillations are also present in  1600  \AA\ light-curves as originally observed by TRACE
(e.g., Krijger et al. 2001), and such periodic behavior
is omnipresent within the observed fields of view.
Moreover, the corresponding light-curves were smoother than those we obtained for ALMA.

The spatially averaged PSDs for the full FoV, cell and network
as well as the corresponding fittings with the function of Equation
\ref{psd} are displayed in Fig. \ref{psdfit1600}. As for the  ALMA
PSDs, we note the existence of a strong peak in the p-mode domain in each
PSD. In addition, the fittings of the PSDs (black lines) with Equation
\ref{psd} (red lines) do a good job, and lead to residuals (blue lines)
which do not exceed $\approx$ 10 $\%$.  

The peak frequency (i.e.,
parameter $a_{3}$ of Equation \ref{fitpsd})
and the relative ($\%$) rms of the p-modes
(i.e., the ratio between
the rms corresponding the Gaussian part of Equation \ref{fitpsd}
as resulting from the application of Equation \ref{rms})
and
the average recorded intensity ($<I>$)
are plotted in panels (c) and (d) of Fig.  \ref{clvalma} for the entire FoV, cell and network. The
corresponding values are tabulated
in Table 2.
From  Fig. \ref{clvalma} we observe that the peak
frequency is $\approx$  3.6 mHz with moderate increase (decrease) towards  
disk center and cell (network) within though the limit
set by the frequency resolution of the obervations. The relative  rms is up to 6\% with a 
weak increase towards disk center (cell, average).

Inspection of the corresponding 304 \AA \, data showed that the light-curves
of the bulk of  individual pixels were rather noisy and it was hard
to infer periodic behavior. On the other hand, only a few pixels per target,
with higher signal-to-noise ratio, showed oscillatory patterns.
Since, (i) we are interested into oscillations characterizing extended, and thus
representative, areas over the observed FoVs, and (ii) we used the AIA data
for mainly supplying context to our ALMA oscillations we refrain from further
discussion of the 304 \AA \, data. Extended temporal sequences as well 
as spatial binning would certainly bring up better the oscillations in 304 \AA \,
something
which, as discussed above, 
 is 
 beyond the scope our study.

\subsection{Time-lag analysis}
We finally performed a cross-correlation analysis of co-spatial AIA 1600 \AA \, and ALMA 3 mm light-curves
in order to determine temporal lags between the two emissions. This would
allow to trace the oscillations between two different layers in the solar atmosphere and
to eventually 
provide an estimate of the difference
between the formation heights of the two emissions. 

\begin{figure}
\centering
\includegraphics[width=0.8\hsize]{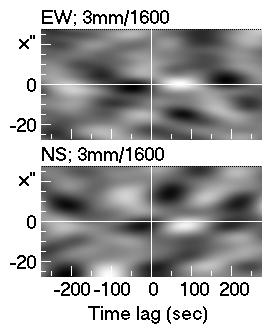}
\caption{2D cross correlation maps of the AIA 1600 \AA \, and ALMA 3 mm
light-curves 
as a function of temporal lag and location along the employed
horizontal (upper panel) and vertical (lower panel) cuts for target 5. 
White (black) correspond to strong positive (negative)
correlation.
The range of values is from -0.22 to 0.25.}
\label{delay}
\end{figure}

Fig. \ref{delay} 
displays the cross-correlation function between the light-curves at 3\,mm and 1600\,\AA, as a function of time lag, 
along two cuts (in the EW and NS direction) through a strong oscillating element in target 5.
It shows strong peaks in the cross-correlation function 
at a lag of $\approx$ 80 s.

For a more detailed study, we computed 
the cross-correlation coefficient 
between
the
AIA 1600 \AA \, and ALMA 3 mm light-curves 
averaged
over 
3\arcsec$\times$ 3\arcsec\ 
macropixels,
so that to decrease noise,
 and for an array of different temporal lags
for all targets.
For further analysis we only kept macropixels with maximum cross-correlation 
coefficients $\ge$ 0.7 so that to ensure strong correlation.
A significant fraction of the total number of macro-pixels satisfied the
above criterion ($\sim 0.25-0.45$). The optimal temporal lag
was calculated from a parabolic fitting of the cross-correlation
function around its peak. 

\begin{figure}
\centering
\includegraphics[width=0.95\hsize]{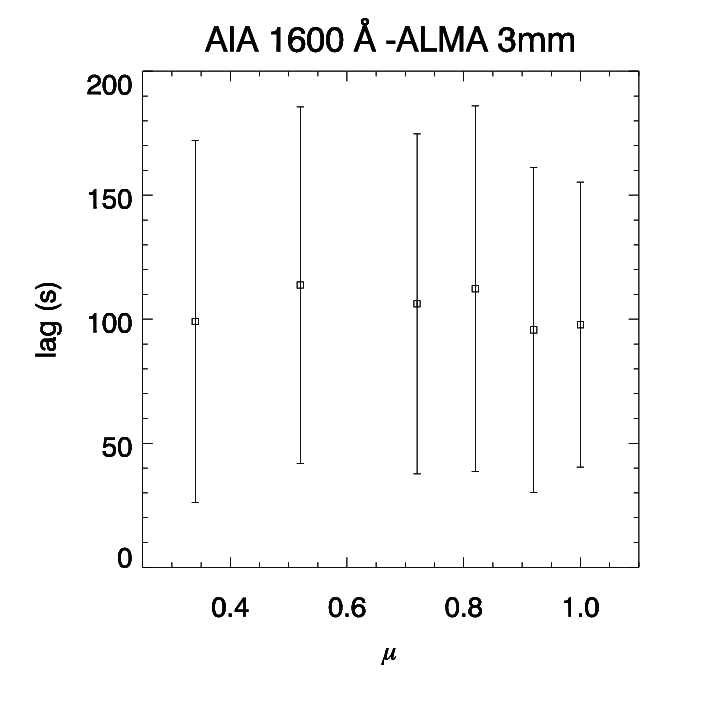}
\caption{Center-to-limb variation of the average lag (squares)
between AIA 1600 \AA \, and ALMA 3 mm
light-curves for pixels corresponding to maximum cross-correlation $\ge$ 0.7.
The vertical error bars specify the standard deviations of the corresponding  
distributions. Positive lag corresponds to AIA 1600 \AA \, proceeding 
ALMA 3 mm.}
\label{clvdelay}
\end{figure}

Our results are shown in Fig. \ref{clvdelay}.
There is no significant center-to-limb variation  of the temporal lag between
AIA 1600 \AA \, and ALMA 3 mm, and its average value is $\approx$ 100 s, therefore
suggesting that  ALMA 3 mm lags AIA 1600 \AA \, by $\approx$ 100 s
; this is slightly larger than the lag suggested by Fig. \ref{delay}, which is within the dispersion of our measurements 
(error bars in Fig. \ref{clvdelay}).
Given now that models and observations show that  AIA 1600 \AA \, forms below ALMA 3 mm
(e.g., Shibasaki et al. 2011; Howe et al. 2012;  
Wedemeyer et al. 2016;
Alissandrakis et al. 2017;  Alissandrakis and Valentino 2019),
our results imply \textit{upward} propagating
waves, consistent with p-mode propagation throughout the chromosphere.
The decreasing relative rms of the p-mode oscillation in going from
1600 \AA \, to ALMA 3 mm is also consistent with this assertion.
We further note that Lindsey \& Roellig (1987) reported a phase delay of 35\degr\ between oscillations at 0.8\,mm and 0.35\,mm in the frequency range of 3-4\,mHz, which corresponds to a time lag of $\sim28$\,s.

The validity of the above result might be questioned 
on the basis of the fact that the measured peak frequencies of the ALMA 3 mm and AIA 1600\,\AA\ oscillations are different,
$\approx$ 4.2 and 3.6\,mHz, respectively; this 
 would
lead to  
a lag of monochromatic oscillations with the same initial phase,
so 
that,  after some time,  
one oscillation would overtake the other. However, 
this possibility 
can
be rather
safely excluded, given the properties of the observed PSDs (e.g., Figures 3 and 4; Tables 1 and 2). Namely,
the difference between the ALMA 3 mm and  AIA 1600\,\AA\ oscillation frequencies is only 0.5 mHz, or about
30\% of the frequency resolution of our observations (i.e., 1.7 mHz).

In addition,
the observed 
oscillations are not monochromatic; they are much broader than the frequency difference,
with FWHM of $\approx$  4.5\,mHz for ALMA 3 mm and somehow
smaller for AIA  1600\,\AA\ in all targets.   
Both these facts imply 
that, within the limitation of the ALMA observations, differences 
between the power spectra of the two data sets are indistinguishable.

Moreover,
if anything, the slightly higher ALMA oscillation frequency compared to AIA 1600,\AA\  
would
eventually lead
to AIA 1600\,\AA\ lagging  ALMA 3 mm, rather preceding it, as our analysis suggests.

We may thus safely conclude that the time lag is real. Before 
interpreting the lag in terms of a height difference, 
a discussion
on the dependence of the 
ALMA and AIA emissions on plasma parameters is warranted.
The fact that the ALMA 3 mm images exhibit similar morphology
as the AIA 1600 \AA\  images (see Fig. 1 of Nindos et al. 2018),
strongly suggests that the two emissions have similar dependencies 
on physical parameters; this is quantified by the high linear correlation coefficients
between the  ALMA 3 mm and AIA 1600\,\AA\ images taking
values in the  range of $\approx$ 0.6-0.8 for all targets.

We now supply a further discussion of the matter based on previous works.
The mm emission, being optically thick, is 
directly related to plasma temperature, as per application of the 
Rayleigh-Jeans law.
For the AIA 1600 \AA \, 
channel  the situation is more complicated. First note that the AIA 1600 \AA \, channel intensities
correspond to a band-pass few hundred\,\AA\ wide around 1600 \AA \, (Boerner et al. 2012).
The emission in this spectral region is dominated by continuum emission via 
bound-free transitions  from  neutral  Si, and from line transitions
(e.g., Vernazza et al. 1981; Fossum and Carlsson 2005), 
and the bound-free absorption coefficient depends on the plasma
temperature. Therefore, one would expect a relationship, 
albeit somehow complex due to radiative transfer
complications, including non-LTE effects, and the non-linear character of the Planck function,
between local plasma temperature and UV contiuum intensities. 
Indeed, detailed 1D non-LTE radiation hydrodynamic simulations
of broad-band acoustic waves in the lower solar atmosphere,
with frequencies covering these of our observations,
showed that there exist heights where the local plasma temperature
is correlated with the corresponding 1600 \AA \, TRACE channel intensity (Fossum and Carlsson 2005).
In addition, 3D non-LTE radiation hydrodynamic simulations of waves in the lower solar atmosphere
by Wedemeyer et al. (2004),
showed a  decent correlation between a temperature  slice  at 500 km, i.e., within
the range of formation heights of 1600 \AA \, discussed below, and 
the corresponding 1600 \AA \, intensity 
(check panels c and h of Figure 2 Wedemeyer et al. (2004)). In summary, 
we conclude that the intensities 
of both analyzed emissions depend primarily on temperature, 
although the details of on the exact dependence 
and its properties are far more obscure 
in the case of 1600\,\AA.

Assuming now that the observed p-mode  oscillations
in  ALMA 3 mm and AIA 1600 \AA \, are  upward-propagating sound waves traveling at  a speed
of $\approx$ 12 $\mathrm{km{s}^{-1}}$  (i.e., corresponding to a  temperature
of 7000 K),
and for the average temporal lag of $\approx$ 100 s between the two
emissions we derived above, we find  a height-separation of 
$\approx$ 
1200  km between  the formation layers 
of ALMA 3 mm and AIA 1600 \AA. Our findings regarding the height-separation between the
 ALMA 3 mm and AIA 1600 \AA \, formation layers
are consistent with related observational and modeling work.
Empirical modeling of the solar atmosphere indicates that
1600 \AA \, forms at a height of $\approx$ 500-750 km 
(Shibasaki et al. 2011), whereas, analysis of TRACE observations of the 1999 Mercury transit
showed a peak height of 500 km and a limb height of 1200 km (Alissandrakis and Valentino 2019).
Radiation-hydrodynamics modeling of a broad-band spectrum 
of acoustic waves
yields a formation height of the 1600 \AA \, channel of TRACE, with similar
wavelength response to the corresponding AIA channel,
of 430 $\pm$ 185 km (Fossum and Carlsson 2005).
Regarding ALMA 3 mm, empirical models show a formation range of $\approx 500-1500$ km
(Wedemeyer et al. 2016) and $\approx 1500-1800$ km (Molnar et al. 2019);
the center-to-limb variation ALMA study of  Alissandrakis et al. (2017) suggest a range 980-1990 km.

\section{Summary and conclusions}

We performed the first study of chromospheric oscillations
in the mm-domain
with ALMA with a resolution of a few arcsec.
Our main findings are summarized as following:
\begin{itemize}
\item p-mode oscillations  at 3\,mm are omnipresent in both network and cell 
with frequencies of $ 4.2 \pm 1.7$\,mHz
(periods $240\pm80$\,s).
\item Oscillation amplitudes of up to  a few hundred K at individual pixels can be found. 
\item The spatially-averaged rms amplitude of oscillations is 
small ($\approx$ 1 $\%$ of the average brightness temperature) and exhibits  a moderate center-to-limb increase.
\item Within the limits of our spectral resolution of 1.7\,mHz, we did not find any significant differences 
in peak frequency between network and cell interior and for different positions from the center of the solar disk to the limb.
\item Our analysis of simultaneous AIA 1600\,\AA\ images gave p-mode oscillations with similar frequencies, 
but stronger by a factor of $\sim6$.
\item ALMA 3 mm lags AIA 1600\,\AA\ by $\sim100$\,s; assuming upward propagating sound waves, 
and asserting that the emission 
intensities depend primarily on temperature in both wavelengths,
this translates to  
a formation height difference of $\approx$ 1200 km.
\end{itemize}

Comparing our findings with 
the lower spatial resolution BIMA observations 
at 3.5 mm of White et al. (2006), we note that these authors found oscillations for both the cell and network
with periods mainly in the domain 210-270 s and 270-330 s, close to the periods
resulting from our analysis of the ALMA oscillations.
However, given
that their  
observing sequences had a duration of 30 min, there were also able to detect longer-period ($>$ 5 minute)
oscillations in the network. This we  could not check with our ALMA observations, given
that
we were observing each target for only 10 minutes.  White et al. (2006) also found rms brightness
temperature amplitudes at
individual pixels in the range 50-150 K,
while the averaged QS time-series
had an rms of 21 K. We found larger values for the oscillation amplitudes 
at individual pixels
(up to a few hundred K), and a larger QS oscillation rms ($\approx$ 55-72 K), which are both
suggesting that chromospheric oscillations in the mm-domain are not fully resolved at a resolution comparable to
the length-scale of the network (i.e., $\approx$ 10\arcsec). Given now that the spatial coherence 
of our ALMA observations has a length-scale comparable to the resolution of our ALMA observations implies
that we are still not fully resolving individual oscillating elements. Higher spatial resolution
ALMA are required in order to investigate this.

Before closing, we note that given the omnipresence  of p-mode oscillations,
and the significant oscillation amplitudes at individual pixels,   
any study of transient chromospheric brightenings needs to
take them into account.

\begin{acknowledgements}
We would like to thank the referee for useful comments/suggestions.
This work makes use of the following ALMA data:
ADS/JAO.ALMA2016.1.00572.S. ALMA is a partnership of ESO (representing its member states),
NSF (USA) and NINS (Japan), together with NRC (Canada) and NSC and ASIAA (Taiwan), and KASI
(Republic of Korea), in cooperation with the Republic of Chile. The Joint ALMA Observatory is
operated by ESO, AUI/NRAO and NAOJ.
\end{acknowledgements}

\end{document}